\documentclass[reprint,
 superscriptaddress,
 bibnotes,
 amsmath,amssymb, aps,
 prl,
floatfix,
]{revtex4-1}

\usepackage{graphicx}
\usepackage[caption=false]{subfig}
\usepackage{dcolumn}
\usepackage{multirow}
\usepackage{tabularx}
\usepackage{bm}
\usepackage{hyperref}
\usepackage{amsmath}
\usepackage{amssymb}
\usepackage{comment}
\usepackage{color}
\usepackage{enumerate}
\usepackage{booktabs}
\usepackage{siunitx}

\newcommand{\GeV}{\text{\,GeV}}

\newcommand{\eV}{\text{~eV}}

\newcommand{\cm}{\, \text{cm}}

\newcommand{\radsi}{\text{~rad} \text{~s}^{-1}}

\newcolumntype{L}[1]{>{\hsize=#1\hsize\raggedright\arraybackslash}X}%
\newcolumntype{R}[1]{>{\hsize=#1\hsize\raggedleft\arraybackslash}X}%
\newcolumntype{C}[1]{>{\hsize=#1\hsize\centering\arraybackslash}X}%

\makeatletter
\newcommand{\thickhline}{%
    \noalign {\ifnum 0=`}\fi \hrule height 2pt
    \futurelet \reserved@a \@xhline
}
\newcolumntype{"}{@{\hskip\tabcolsep\vrule width 2pt\hskip\tabcolsep}}
\makeatother

\newcommand{\Eq}[1]{Eq.~(\ref{#1})}

\newcommand{\Fig}[1]{Fig.~\ref{#1}}

\definecolor{deepblue}{RGB}{20,20,160}

\begin{document}

\title{
Search for ultralight scalar dark matter with atomic spectrosopy
}

\author{Ken Van Tilburg}\email{kenvt@stanford.edu}\affiliation{Stanford Institute for Theoretical Physics, Stanford University, Stanford, CA 94305, USA}
\author{Nathan Leefer}\email{naleefer@uni-mainz.de}\affiliation{Helmholtz Institut Mainz, 55128 Mainz, Germany}
\author{Lykourgos Bougas}\email{lybougas@uni-mainz.de}\affiliation{Helmholtz Institut Mainz, 55128 Mainz, Germany}
\author{Dmitry Budker}\email{budker@uni-mainz.de}\affiliation{Helmholtz Institut Mainz, 55128 Mainz, Germany}\affiliation{Institut f\"ur Physik, Johannes Gutenberg Universit\"at-Mainz, 55128 Mainz, Germany}\affiliation{University of California at Berkeley, Berkeley, CA 94720, USA}

\date{\today}

\begin{abstract}
We report new limits on ultralight scalar dark matter (DM) with dilaton-like couplings to photons that can induce oscillations in the fine-structure constant $\alpha$. Atomic dysprosium exhibits an electronic structure with two nearly degenerate levels whose energy splitting is sensitive to changes in $\alpha$.
Spectroscopy data for two isotopes of dysprosium over a two-year span is analyzed for coherent oscillations with angular frequencies below $1 \radsi$.
No signal consistent with a DM coupling is identified, leading to new constraints on dilaton-like photon couplings over a wide mass range.
Under the assumption that the scalar field comprises all of the DM, our limits on the coupling exceed those from equivalence-principle tests by up to 4 orders of magnitude for masses below $3 \cdot 10^{-18} \text{~eV}$.
Excess oscillatory power, inconsistent with fine-structure variation, is detected in a control channel, and is likely due to a systematic effect. 
Our atomic spectroscopy limits on DM are the first of their kind, and leave substantial room for improvement with state-of-the-art atomic clocks.
\end{abstract}

\maketitle



Dark matter (DM) makes up the majority of matter density in  our Universe.  Its ubiquitous abundance can be measured through its gravitational influence, but little is known about the microphysical properties of the DM particle(s), such as the mass, spin, and any nongravitational interactions.
If the DM is bosonic rather than fermionic, it can have a sub-eV mass and such high occupation numbers that it acts more like a classical wave---with frequency equal to its mass---than a particle. Light bosonic dark matter has a natural production mechanism, namely early-universe misalignment of the field relative to the minimum of its potential~\cite{Preskill:1982cy,Dine:1982ah,Abbott:1982af}. Several motivated candidates in this category exist in the literature, most notably the QCD axion~\cite{Peccei:1977hh,Weinberg:1977ma,Wilczek:1977pj} and other axion-like particles, which, as parity-odd pseudo-Nambu-Goldstone bosons (PNGBs) of compact symmetry groups, primarily have derivative interactions with matter \cite{Graham:2013gfa}. Light, parity-even bosons may also arise as PNGBs of noncompact groups such as those of scale, conformal, or shift symmetries, the most famous examples of which are dilatons \cite{Taylor:1988nw,Ellis:1987qs,Damour:1994zq}. Small explicit breakings of these symmetries may induce nonderivative operators for the scalar fields, such as mass terms and higher-dimensional operators coupling them to matter.


We focus on ultralight scalar fields $\phi$ with couplings to the (square of the) electromagnetic field tensor $F_{\mu \nu}$:
\begin{align}
\mathcal{L} \supset \frac{1}{2} (\partial_\mu \phi)^2 -\frac{1}{2}m_\phi^2 \phi^2 + \frac{-1 + d_e \kappa \phi}{4e^2}F_{\mu\nu}F^{\mu\nu}, \label{eq:lagrangian}
\end{align}
where $\kappa \equiv \sqrt{4\pi G_N}$, $G_N$ is Newton's constant, and $e \approx 0.303$ is the electromagnetic gauge coupling (we use units in which $\hbar = c = 1$).  The interaction is normalized such that $d_e = 1$ yields an attractive force of gravitational strength between electromagnetic energy densities at distances smaller than the inverse mass ($r \lesssim m_\phi^{-1}$) through scalar $\phi$ exchange~\cite{Damour:2010rp}. Couplings $d_e \ll 1$ generically arise if quantum gravity effects weakly break an underlying global symmetry of $\phi$ near the Planck scale $\sim G_N^{-1/2}$~\cite{Kallosh:1995hi}. Quantum corrections proportional to $d_e^2$ and naturalness considerations together suggest a minimum mass-squared for $\phi$. However, given the existing hierarchy problems of the Standard Model, we remain agnostic to this issue and consider the full $m_\phi$--$d_e$ parameter space (see Ref.~\cite{Arvanitaki:2014faa} for more discussion). 

Equivalence principle (EP) tests such as the E\"ot-Wash experiment~\cite{Schlamminger:2007ht} and Lunar Laser Ranging~\cite{Williams:2004qba} constrain the coupling $d_e$ to be much less than unity: $|d_e| \lesssim 3.6 \cdot 10^{-4}$ at 95\% confidence level (CL) for $m_\phi \lesssim 3 \cdot 10^{-14}\eV$. This limit will likely be improved to $|d_e| \lesssim 7.8 \cdot 10^{-5}$ with atom-interferometry techniques~\cite{Dimopoulos:2006nk}. The EP-violating force that $\phi$ mediates scales as $|d_e|^2$, making vast improvements to these limits challenging. 

Light scalar fields do not behave as perfect cold DM on short length scales, where their density perturbations have a nonzero sound speed. 
For $m_\phi \lesssim 10^{-22}\eV$, they would have inhibited cosmological structure growth~\cite{Marsh:2013ywa,Hlozek:2014lca,Bozek:2014uqa} in conflict with observations, though these bounds disappear if $\phi$ only makes up a small fraction of the dark matter. 
Because of its effect on structure formation, light scalar dark matter in the $10^{-24}$--$10^{-20}\eV$ range has been proposed~\cite{Hu:2000ke,Marsh:2015wka} to solve several long-standing astrophysical puzzles, such as the core-cusp, missing satellite, and too-big-to-fail problems~\cite{Weinberg:2013aya}. DM self-interactions such as cubic or quartic potential terms likewise produce pressure contributions~\cite{Turner:1983he}; if they are sufficiently small in the early universe, they can be neglected in the present era~\cite{Arvanitaki:2014faa}.


In Ref.~\cite{Arvanitaki:2014faa}, it was pointed out that the interaction in \Eq{eq:lagrangian} leads to a fractional oscillation in the fine-structure constant if $\phi$ comprises the dark matter:
\begin{align}
\alpha(t) \simeq \alpha \left[1 + d_e \kappa \phi_0 \cos(m_\phi t + \delta) \right], \label{eq:alphavar}
\end{align}
with $\alpha \equiv e^2/4\pi$ and $\delta$ an arbitrary phase. The amplitude $\phi_0$ depends on the ambient dark matter energy density $\rho_\text{DM} \approx 0.3 \GeV \cm^{-3}$ as:
\begin{align}
 \kappa \phi_0 \simeq \kappa \frac{\sqrt{2 \rho_\text{DM}}}{m_\phi} \approx 6 \cdot 10^{-16} \left( \frac{10^{-15}\eV}{m_\phi} \right).\label{eq:amplitude}
\end{align}
The amplitude $\phi_0$ scales as the square root of the energy density, so even if $\phi$ makes up a tiny fraction of the DM energy density, the oscillation in \Eq{eq:alphavar} may still be detectable. The field oscillation has an angular frequency equal to $m_\phi$, and a coherence time of order $2\pi / m_\phi v^2 $, with $v \sim 10^{-3}$ the velocity dispersion of the DM in our Galaxy~\cite{calcaxion}. The amplitude in \Eq{eq:amplitude} scales inversely with the frequency of the oscillation, motivating measurements of this effect at low frequencies. Our measurements are sensitive to angular frequencies below $1 \radsi$, or DM masses $m_\phi \lesssim 10^{-15}\eV$, for which we can take the DM oscillations to be coherent on the time scales of our experiment.
At angular frequencies higher than $1 \radsi$, experiments with low noise floors such as gravitational-wave observatories and other resonant detectors may also reach sensitivity beyond existing bounds~\cite{Arvanitaki:2014faa}, despite the smaller oscillation amplitude.

Transition energies in atoms are convenient observables to look for low-frequency signals of changes in masses and couplings, and have been used to look for linear drifts in $\alpha$~\cite{Leefer2013a,Rosenband28032008,PhysRevLett.98.070801,PhysRevLett.113.210801,Huntemannn2014,Guena2012,Tobar2013}. Atoms are exactly reproducible systems across long time scales, their electronic structure solely determined by masses and couplings of the Standard Model. While experimental imperfections can lead to fluctuations in observed transition frequencies, recent advances in optical metrology techniques have allowed for the determination of transition frequencies with stabilities and accuracies approaching $\delta{\nu}/\nu \sim 10^{-18}$~\cite{Hinkley13092013,Bloom:2013uoa,Huntemann:2012zz,Oskay:2006zz,PhysRevLett.104.070802}. 

In this work, we perform a spectroscopic analysis in two isotopes of dysprosium to search for the coupling in \Eq{eq:lagrangian} through the effect of \Eq{eq:alphavar}. In the remainder of this Letter, we describe the atomic level structure of dysprosium and the experimental setup, followed by an account of the data analysis. No statistically significant oscillation consistent with $\alpha$ variation is observed; new limits are placed on couplings of DM lighter than $10^{-15}$ eV. 

\begin{figure}[t]
\includegraphics[width = 0.99\columnwidth]{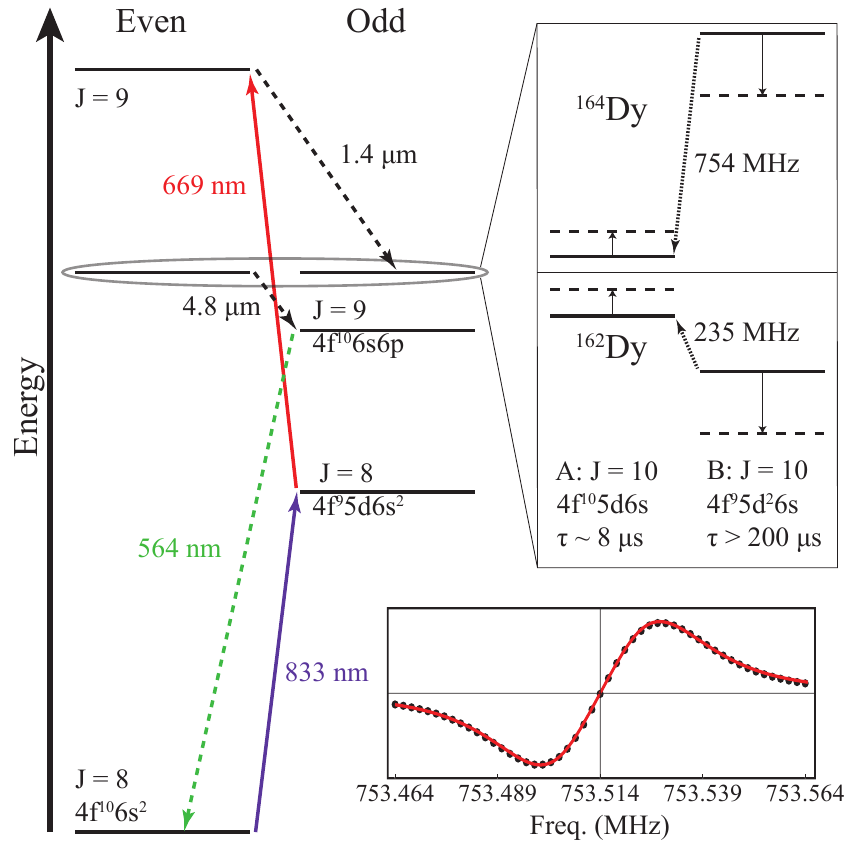}
\caption{Energy level diagram of dysprosium showing the nearly-degenerate pair of states $A$ and $B$, and the transitions used for state preparation and fluorescence detection. The inset shows the $A$ and $B$ states for $^{164}$Dy (754 MHz) and $^{162}$Dy (235 MHz), and their response to a positive change in $\alpha$: the transition frequency increases in $^{162}$Dy and decreases in $^{164}$Dy. The bottom-right graph displays a lineshape derivative for the 754 MHz transition in $^{164}$Dy.
}\label{fig:levels}
\end{figure}

Dysprosium (Dy) is a rare-earth element with nuclear charge $Z = 66$ and 7 stable isotopes, of which we use those with atomic masses $A = 162~\&~164$.
A large number of valence electrons endows Dy with a complex energy level structure, including a nearly degenerate pair of opposite-parity states, denoted $A$ and $B$ for the even and odd states, respectively~\cite{Dzuba1986}.
Spectroscopy of the radio-frequency (rf) electric-dipole transition between $A$ and $B$ revealed that their energy splitting corresponds to frequencies less than $2000$ MHz, and that in some isotopes, $A$ is the more energetic state ($^{162}$Dy), while in others ($^{164}$Dy), $B$ has the higher energy \cite{Budker1994}. 
Figure~\ref{fig:levels} depicts the relevant energy level structure, with a focus on $^{162}$Dy and $^{164}$Dy.

The energy splitting between the nearly degenerate pair is extremely sensitive to variation of the fine-structure constant. A change in $\alpha$ yields a frequency shift $\delta \nu = \pm \nu_{\delta \alpha} \,\delta\alpha/\alpha$ with $\nu_{\delta \alpha} \approx 2 \cdot 10^{15}\text{\,Hz}$~\cite{Dzuba2008}.
By comparison, optical frequency measurements with trapped atoms and ions have similar absolute sensitivities to variation in $\alpha$, but the near-degeneracy of the $A$ and $B$ levels
relaxes requirements on the fractional accuracy and stability of the frequency reference.
Spectroscopy in Dy resulted in one of the most stringent constraints on a present-day, linear variation of $\alpha$ at the level of $|\Delta \alpha/\alpha| \lesssim 10^{-16}$~yr$^{-1}$, the best limit on a possible coupling of $\alpha$ to gravitational potential~\cite{Leefer2013a}, and stringent limits on violation of Lorentz invariance for electrons~\cite{Leefer:2013yaa}. 


\begin{figure}[t]
\includegraphics[width = 0.99\columnwidth]{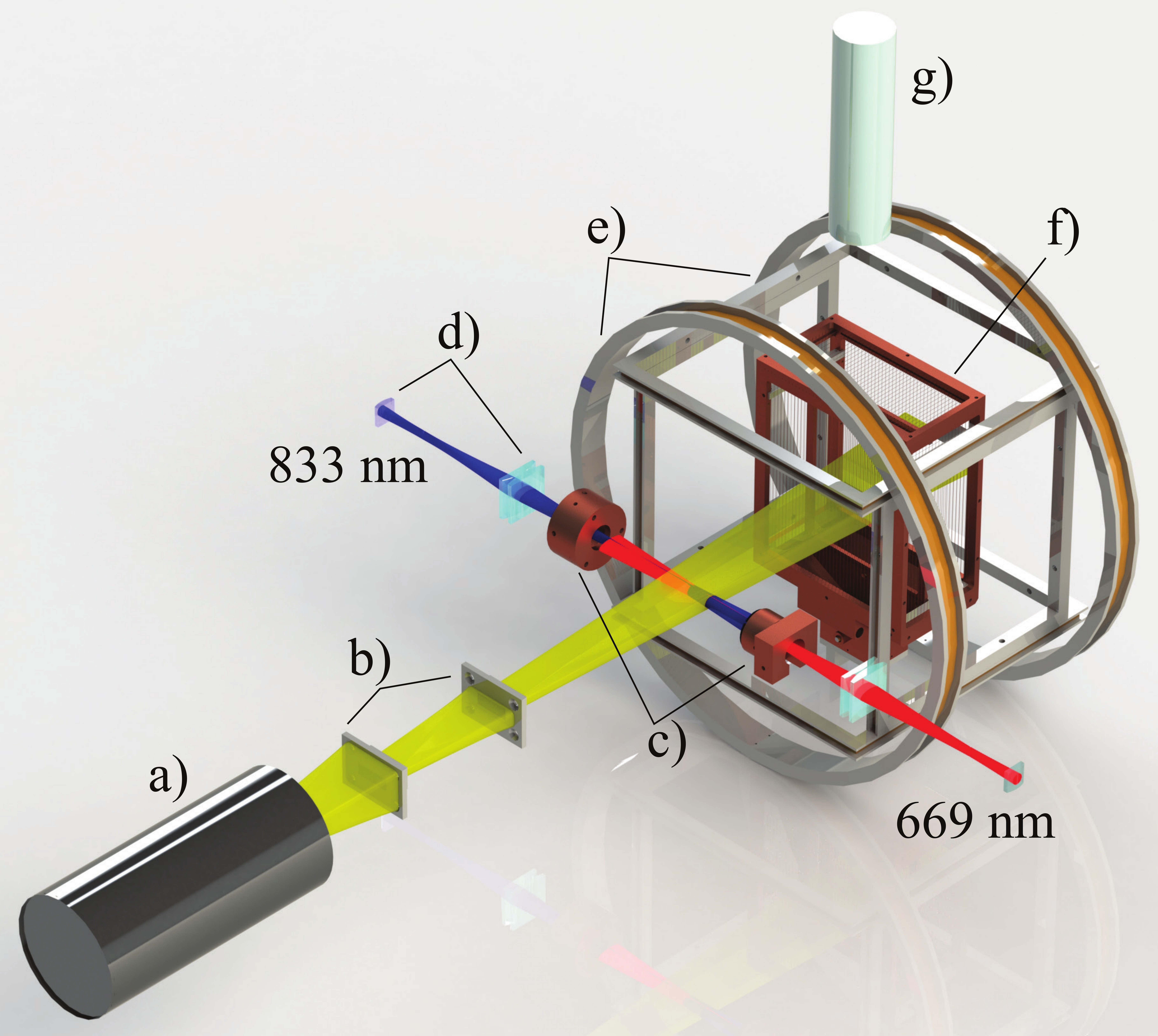}
\caption{Simplified apparatus layout. \textbf{a)} Thermal beam of dysprosium atoms an oven heated to 1400 K. \textbf{b)} Skimmers collimate the beam. \textbf{c)} Linear polarizers suppress residual ellipticipy due to birefringence of vacuum chamber windows. \textbf{d)} Cylindrical lenses match the divergence of the atom and laser beams, for efficient population transfer~\cite{Nguyen2000}. \textbf{e)} Magnetic field coils compensate stray magnetic fields. \textbf{f)} Radio-frequency field interaction region. \textbf{g)} Photomultiplier tube for fluorescence detection.}\label{fig:apparatus}
\end{figure}

A beam of dysprosium atoms is generated by an oven operating at $1400$ K, and prepared in state $B$ by two laser excitations followed by a spontaneous decay, as shown in \Fig{fig:levels}. The lifetime of state $B$ is long enough that it can be considered metastable in this experiment~\cite{Budker1994}. An rf electric field, whose frequency is compared with a cesium reference and a GPS-stabilized rubidium oscillator, then excites atoms to state $A$, which decays to the ground state through several channels, one of them emitting a 564-nm photon. This fluorescence is detected with a photomultiplier tube (PMT), allowing the resonant frequency to be determined by maximizing fluorescence with respect to the electric-field frequency. A simplified illustration of our setup is shown in \Fig{fig:apparatus}, and more details can be found in Refs.~\cite{Cingoz2008,Weber2013}. 
The optimal statistical precision is $\sigma_\nu = \gamma/(2\pi\sqrt{\dot{N} \tau})$, where $\gamma/(2\pi) \approx 20$~kHz is the natural linewidth of the transition (see bottom panel of \Fig{fig:levels}), $\dot{N}$ is the number of fluorescent photons per unit time, and $\tau$ is the total integration time. An estimate of $\dot{N} = 10^9 \text{~s}^{-1}$ gives a statistical measurement precision of $\sigma_\nu \sqrt{\tau} = 0.6 \text{~Hz}/\sqrt{\text{Hz}}$. In the actual experiment, the statistical background is dominated by leakage into the PMT of blackbody radiation and scattered light from the 669-nm excitation laser, limiting the statistical precision for 
$^{162}$Dy and $^{164}$Dy to $4 \text{~Hz}/\sqrt{\text{Hz}}$. This corresponds to a sensitivity of $\delta \alpha/\alpha \approx 2 \cdot 10^{-15}$ after one second of integration, and improves as $1/\sqrt{\tau}$.

Our data consist of transition frequency measurements on $^{162}$Dy and $^{164}$Dy---denoted by $\nu_{162} \approx 235$~MHz and $\nu_{164}\approx 754$~MHz, respectively---recorded at different times $t_i$.
The energy hierarchy of states $A$ and $B$ is reversed between these isotopes, which means that the response of their transition frequencies to a variation of $\alpha$ is opposite in sign,
shown diagramatically in \Fig{fig:levels}. 
An increase in $\alpha$ would \emph{increase} the transition frequency between the $A$ and $B$ states in $^{162}$Dy, but \emph{decrease} it in $^{164}$Dy.
With measurements in both isotopes, two types of data sets can be constructed: one ``out-of-phase" signal data set maximally sensitive to $\alpha$ variation, the other an ``in-phase" control channel data combination which can be checked for systematic effects. Because each isotope's transition frequency was not observed simultaneously, any systematic effects in the control data need to be subtracted to account for potential spectral leakage of these effects into the signal data. 


Measurements were performed on 10 different days over a two-year period (2010--2012), as in Ref.~\cite{Leefer2013a}. We split up the data into two different sets, one long-term data set (hereafter ``LT") of combined measurements on the first 9 measurement days, and a short-term data set (``ST") of the last day of measurements. In the LT data set, the measurements of $\nu_{162}$ and $\nu_{164}$ are averaged over the course of each of the 9 days, such that each day-averaged measurement is dominated by a systematic error of 0.48 Hz (1 Hz on the first day, due to a different experimental configuration). After subtracting the overall mean in both LT sets, we define the LT signal data set as $\nu_\text{signal}(t_i)\equiv \nu_{162}(t_i) - \nu_{164}(t_i)$, sensitive to variations of $\alpha$. Adding the frequencies from each isotope \emph{in phase} defines the LT control data set $\nu_\text{control}(t_i)\equiv \nu_{162}(t_i) + \nu_{164}(t_i)$. The ST data set comprises 2303 measurements of $\nu_{162}$ and $\nu_{164}$ taken over a span of $14.5\text{ h}$ on 19 October 2012. Data from each isotope are mean-subtracted, and combined into the ST signal time series data as $\nu_\text{signal}(\lbrace t_i,t_j \rbrace)  =  \nu_{162}(\lbrace t_i \rbrace)  -  \nu_{164}(\lbrace t_j \rbrace)$. Similar to the LT data combination, we also construct a ST control data set $\nu_\text{control}(\lbrace t_i,t_j \rbrace)$. 


Our aim is to search for harmonic variations in the dysprosium transition frequency data. A signal or limit at an angular frequency $\omega$ can then be converted into a signal or limit for scalar dark matter with coupling $d_e$ and mass $m_\phi \simeq \omega$. We perform linear least-squares analysis (LLA) on $\nu_\text{control}(t_i)$ and $\nu_\text{signal}(t_i)$ with waveforms of type $\nu=\nu_0 \cos(\omega t + \varphi) + \nu_\text{c}$ and $\nu_\text{c}$ a constant offset \footnote{A common way to search for harmonics in time series data---in this case measurements $\nu (t_i)$---is the discrete Fourier transform (DFT). However, our data is unevenly sampled ($t_{i+1}-t_{i}$ is different for each $i$), a scenario in which a DFT is ill-suited.}. The highest analyzed angular frequency $\omega_\text{max}$ is taken to be $4\pi/\Delta t_\text{min}$, with $t_\text{min}$ the shortest time between measurements. LLA can also capture variations at angular frequencies smaller than the inverse time span of the data set $T$, when the waveform becomes essentially a frequency drift (near a node) or a quadratic frequency change (near an antinode).

\begin{figure}[t]
\includegraphics[width = 0.99 \columnwidth]{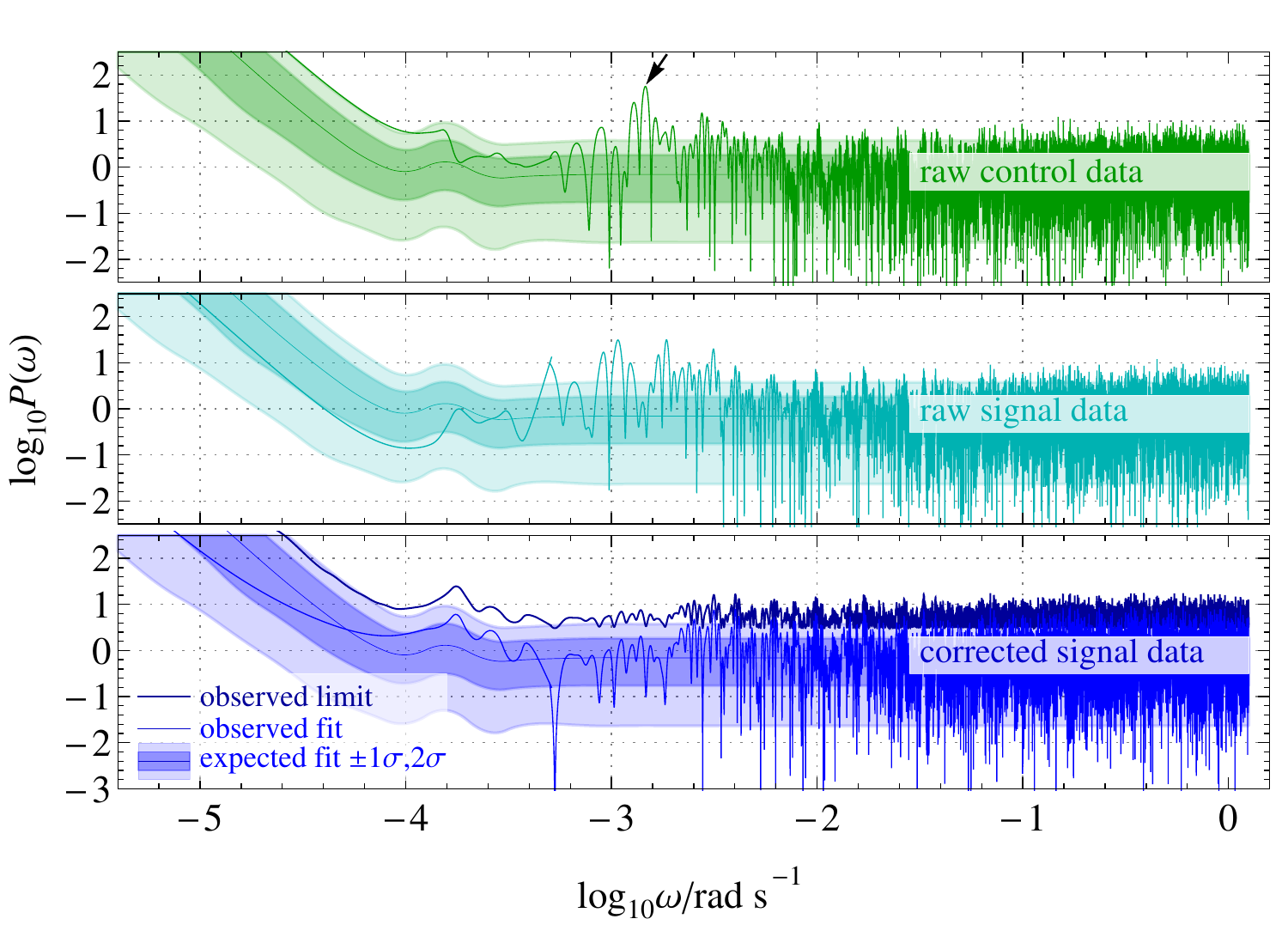}
\caption{Power spectra of the best-fit waveforms for three combinations of the ST data. The thin line and the dark and light bands correspond to the expected median and 68\%- \& 95\%-CL intervals for the best fit. The raw control data set shows excess power at angular frequency $\omega \approx 1.4 \cdot 10^{-3}\radsi$ (indicated by the arrow), which leaks into the raw signal data in nearby frequency bands. After fitting out this effect, the power in both data combinations is consistent with white noise, shown for the signal data in the bottom panel. The 95\% CL upper limit is depicted by a thick, dark blue line above the corrected best-fit signal power.
}\label{fig:power}
\end{figure}

From the best-fit LLA waveform, we construct a normalized power spectrum: 
\begin{align}
P(\omega) \equiv \frac{N_0}{4 \sigma^2_{ \nu}}\nu_0^2,
\end{align}
where $\sigma^2_{\nu}$ is the average variance of $\nu(t_i)$, and $N_0$ is the number of data points.
For high angular frequencies $2\pi/T \lesssim \omega < \omega_\text{max}$, the LLA power spectrum reduces to the modified periodogram of Ref.~\cite{Scargle:1982bw}, and has simple statistical properties for white noise with equal uncertainties on each data point (a good first-order approximation for both of our data sets). In this limit, the power $P(\omega)$ at any frequency has a simple cumulative distribution function (CDF) which is $\omega$-independent: $\text{Prob}\lbrace P(\omega)< P \rbrace = 1-e^{-P}$. It follows that the expected median, and the 68\% and 95\% confidence intervals are $0.69$, $[0.17,1.8]$, and $[0.023,3.8]$, respectively.
At intermediate and low angular frequencies $\omega \lesssim 2\pi/T$, the statistical behavior of the power is quite complicated, and we simulated its expected distribution with Monte Carlo (MC) using the same  errors as on the data. 

We set a 95\% CL upper bound on the power by assuming the observed best-fit oscillation is a true signal, adding white noise via MC many times, and taking the 95th-percentile value of the combined power. At high frequencies, an approximate, analytic expression for this value can be obtained~\cite{1975ApJS...29..285G}.
In the presence of a real signal with power $P_\text{s}$, the CDF of the periodogram is modified to $\text{Prob}(P(\omega) < P \, | \,  P_\text{s}) = 1-e^{-(P+P_\text{s})}\phi(P,P_\text{s})$,
where $\phi(P,P_s) \equiv \sum_{m=0}^{\infty} \sum_{k=0}^{m}P^k P_\text{s}^m / k! m!$.
Given an observed power $P_\text{s}$, we define our 95\%-CL upper limit $P_\text{lim}$ by $0.95 \approx \text{Prob}(P(\omega)<P_\text{lim} \, | \, P_\text{s})$. At very low angular frequencies $\omega \ll 2\pi/T$, where the periodogram statistics are invalid, the expected limit on the power (amplitude) scales as $\omega^{-4}$ ($\omega^{-2}$) because one cannot exclude being near an antinode of an oscillation.

In \Fig{fig:power}, we plot the results of this analysis for the ST data set.  Excess power is observed in the control data---sensitive only to in-phase fluctuations of $\nu_{162}$ and $\nu_{164}$---at high significance: $P(1.4 \cdot 10^{-3}\radsi) \approx 57$. We could not conclusively identify the source of this monochromatic variation, observed independently in both isotopes with approximately equal amplitude, phase, and period of $4.3\cdot 10^3 \text{ s}$. Its best-fit amplitude of $\nu_0 \approx 0.78 \text{ Hz}$ is comparable to the magnitude of a systematic error due to unstable electronic offsets~\cite{Leefer2013a}, while its period may point to room temperature fluctuations. Variations due to reference clock instabilities are excluded because they would produce a larger effect in $^{164}$Dy than in $^{162}$Dy. Upon subtracting the best-fit waveform to the variation in the control data, which partially leaked into the signal data in neighboring frequency bands, the ST corrected power spectra were compatible with white-noise fluctuations. For the raw LT data, the observed power spectra in both the control and signal data were also consistent with white noise.

\begin{figure}[t]
\includegraphics[width = 0.99 \columnwidth]{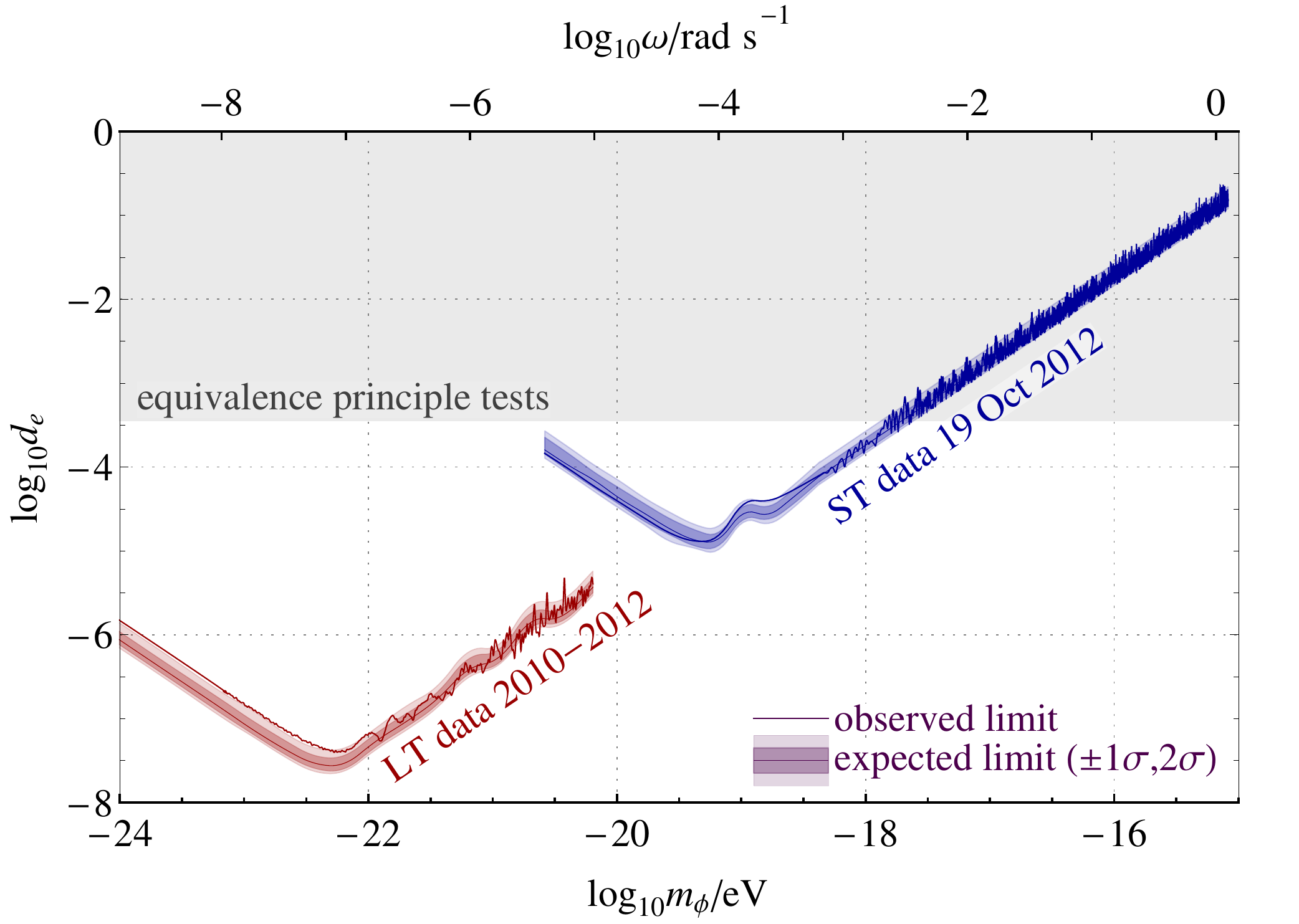}
\caption{Upper limits at 95\% CL on the coupling $|d_e|$ to photons as a function of dark matter mass $m_\phi$. Shaded bands correspond to 68\%- and 95\%-CL intervals around the median expected limit (thin line); thick lines depict the observed upper limits. Parameter space excluded at 95\% CL by EP-violating force tests is shown in gray.
}\label{fig:limit}
\end{figure}

Assuming the field $\phi$ comprises all of the DM, power values can be translated to absolute values of the scalar coupling to photons via:
\begin{align}
|d_e(m_\phi)|\simeq \frac{1}{ \kappa \phi_0(m_\phi) } \left[ \frac{4 \sigma_{ \nu}^2}{N_0 \nu_{\delta \alpha}^2}  P(m_\phi) \right]^{1/2}, \label{eq:delim}
\end{align}
where $\nu_{\delta \alpha} \approx 2 \cdot 10^{15} \text{ Hz}$ is the aforementioned sensitivity coefficient to fractional variations of $\alpha$. Note that any bound on $|d_e|$ scales only as the square root of the energy density in the $\phi$ field via \Eq{eq:amplitude}. Figure~\ref{fig:limit} shows the 95\%-CL upper limit on $|d_e|$ as a function of mass $m_\phi \simeq \omega$ for both data sets. In particular, the observed limit from the ST data is the limit curve of \Fig{fig:power} transformed with \Eq{eq:delim}. For masses $m_\phi \lesssim 3 \cdot 10^{-18}\eV$, our ST limits on $|d_e|$ exceed those set by EP tests.  In our most sensitive mass window---$m_\phi \sim 10^{-22} \text{ eV}$, which coincides with the astrophysically motivated mass range~\cite{Marsh:2015wka}---we exclude DM couplings down to $d_e \approx 4.2 \cdot 10^{-8}$ times gravitational strength.


We have presented the best limit on neutral scalar dark matter coupling to photons for masses below $3 \cdot 10^{-18 } \eV$.  It is also the first limit of its kind---exploiting the response of atomic transition energies to tiny fractional oscillations of the fine-structure constant---and we hope many similar searches will follow. Optical clock systems will likely improve on our limit of $|d_e|$ with sufficient data. Combined with existing microwave and future nuclear clocks, one can also achieve sensitivity to fractional variations in the electron, quark, and proton masses, expanding the potential discovery reach of light scalar dark matter to other couplings~\cite{Arvanitaki:2014faa}.
Small-scale precision experiments can thus search for cosmic dark matter with feeble interactions generated at the highest energy scales.

\acknowledgments{
We thank Asimina Arvanitaki for collaboration in the initial stages of this work, and Masha Baryakhtar, Savas Dimopoulos, Junwu Huang, Xinlu Huang, and Tim Wiser for valuable discussions. NL was supported by a Marie Curie International Incoming Fellowship within the 7th European Community Framework Programme.
}


\bibliographystyle{apsrev4-1-etal}
\bibliography{dylaton}

\end{document}